\documentclass{article}

\usepackage[superscript]{cite}
\usepackage{lineno}
\usepackage{graphicx}
\usepackage{float}
\usepackage[labelfont={bf}, labelsep = colon]{caption}

\DeclareCaptionLabelSeparator{sep}{ $|$ }
\usepackage{color}
\usepackage{newtxtext,newtxmath,amsmath}
\usepackage{multirow}
\usepackage{authblk}
\usepackage[colorlinks,linkcolor=blue, citecolor=blue, urlcolor=blue]{hyperref}

\begin{document}

\begin{center}
\Large \textbf{Measurement of the mass difference and the binding energy of the hypertriton and antihypertriton}
\end{center}

\textbf{The STAR Collaboration}
\\

\textbf{According to the CPT theorem, which states that the combined operation of charge conjugation, parity transformation and time reversal must be conserved, particles and their antiparticles should have the same mass and lifetime but opposite charge and magnetic moment. Here, we test CPT symmetry in a nucleus containing a strange quark, more specifically in the hypertriton. This hypernucleus is the lightest one yet discovered and consists of a proton, a neutron, and a $\Lambda$ hyperon. With data recorded by the STAR detector{\cite{TPC,HFT,TOF}} at the Relativistic Heavy Ion Collider, we measure the $\Lambda$ hyperon binding energy $B_{\Lambda}$ for the hypertriton, and find that it differs from the widely used value{\cite{B_1973}} and from predictions{\cite{2019_weak, 1995_weak, 2002_weak, 2014_weak}}, where the hypertriton is treated as a weakly bound system. Our results place stringent constraints on the hyperon-nucleon interaction{\cite{Hammer2002, STAR-antiH3L}}, and have implications for understanding neutron star interiors, where strange matter may be present{\cite{Chatterjee2016}}. A precise comparison of the masses of the hypertriton and the antihypertriton allows us to test CPT symmetry in a nucleus with strangeness for the first time, and we observe no deviation from the expected exact symmetry.}

The CPT theorem holds that all processes must exactly conserve the combined operation of C (charge conjugation, which interchanges a particle with its antiparticle), P (parity, which reverses the direction of all spatial axes), and T (time reversal). No CPT violation has ever been observed {\cite{PDG, massdiff_ALICE}}. Qualitatively different tests of CPT symmetry are a continuing priority for fundamental physics, as are revisitations of past tests with improved accuracy. While CPT invariance has been verified to a precision of $10^{-19}$ in the strange quark sector for kaons{\cite{PDG}}, we present here the first test of CPT symmetry in a nucleus (multi-baryon cluster) having strangeness content. Similar to recent CPT tests{\cite{massdiff_CMS_2017, massdiff_ATLAS_2014, massdiff_Omega_1998}} on parameters of the Standard Model Extension{\cite{SME_2019, SME_1997}}, the mass difference between hypertriton and antihypertriton is directly constructed from the Lorentz invariant product of the four-momenta of their weak-decay daughters.

Hypernuclei are natural hyperon-baryon correlation systems, and provide direct access to the hyperon-nucleon ($YN$) interaction through  measurements of the binding energy $B_\Lambda$ in a hypernucleus{\cite{binding_2013PRD}}. However, in a half-century of research, the creation of the hypertriton and precise measurement of its properties have proven difficult, in contrast to heavier hypernuclei produced via a kaon beam incident on a nuclear target. Early measurements of the hypertriton $B_\Lambda$ are consistent with zero and span a wide range characterized by a full width at half-maximum of 2.1 MeV{\cite{Achenbach2017}}. Modern facilities now permit an improved understanding of the $YN$ interaction, via improved measurements of hyperon binding in hypernuclei, and through new hypertriton lifetime measurements{\cite{ALICE-H3L-lifetime, STAR-H3L-lifetime}}. Progress in understanding the $YN$ interaction and the equation of state (EOS) of hypernuclear matter has implications for understanding neutron star properties. Inclusion of hyperons in the cores of neutron stars softens the equation of state, and thus reduces the stellar masses{\cite{Chatterjee2016,Ypuzzle}}. In model calculations, the maximum mass of the neutron star depends on the assumed $\Lambda NN$ interaction which is directly related to the $\Lambda$ binding energy in hypernuclei{\cite{Ypuzzle, NS_new}}. A precise binding energy measurement of this simplest hypernucleus together with other light hypernuclei will also help us understand the few-body system and the strong interaction involving hyperons{\cite{overbinding_2018PRL}}.

Nuclear collisions at ultrarelativistic energies, such as those studied at the Relativistic Heavy Ion Collider (RHIC), create a hot and dense phase of matter containing approximately equal numbers of quarks and antiquarks. In this phase, called the quark-gluon plasma (QGP), quarks are free to move throughout the volume of the nuclear collision region. The QGP persists for only a few times $10^{-23}$ seconds, then cools and transitions into a lower temperature phase comprised of mesons, baryons and antibaryons, including the occasional antinucleus or antihypernucleus {\cite{STAR-antiH3L,STAR-anti}}. Thus these collisions offer an ideal laboratory to explore fundamental physics involving nuclei, hypernuclei, and their antimatter partners.

In this letter, we present two measurements from gold-gold collisions at a center-of-mass energy per nucleon pair of $\sqrt{s_{NN}} = 200$ GeV: the relative mass difference between $\rm^3_\Lambda H$ (the hypertriton) and $\rm^3_{\bar{\Lambda}}\overline{H}$ (the antihypertriton), as well as the $\Lambda$ hyperon binding energy for $\rm^3_\Lambda H$ and $\rm^3_{\bar{\Lambda}}\overline{H}$. The $\Lambda$ binding energy of $\rm^3_\Lambda H$ is defined as $B_\Lambda = (m_d + m_\Lambda - m_{^3_\Lambda{\rm H}})c^2$, where $m_d$, $m_\Lambda$, $m_{^3_\Lambda{\rm H}}$ are the deuteron mass taken from the CODATA{\cite{CODATA}}, the $\Lambda$ hyperon mass published by the Particle Data Group (PDG){\cite{PDG}}, and the $^3_\Lambda$H mass reported in this letter, and $c$ is the speed of light.  The main detectors used in this analysis are the Solenoidal Tracker At RHIC (STAR) Time Projection Chamber (TPC){\cite{TPC}} and the Heavy Flavor Tracker (HFT){\cite{HFT}} for high-precision tracking, and the TPC and the Time Of Flight detector (TOF){\cite{TOF}} for charged particle identification. The TPC and HFT are immersed in a solenoidal magnetic field of 0.5 T parallel to the beam direction, and are used for charged particle tracking in three dimensions.  The HFT includes three subsystems: Pixel (PXL), which consists of two cylindrical layers at radii 2.8 and 8 cm from the beam, the Intermediate Silicon Tracker (IST) at a radius of 14 cm, and the Silicon Strip Detector (SSD) at a radius of 22 cm. The spatial resolution of the HFT{\cite{HFT}} is better than 30 $\mu$m for tracks with a momentum of 1 GeV/c. The mean energy loss per unit track length ($\langle dE/dx\rangle$) in the TPC gas and the speed ($\beta$) determined from TOF measurements are used to identify particles. The $\langle dE/dx\rangle$ resolution{\cite{TPC}} is 7.5\% and the TOF timing resolution{\cite{TOF}} is 95 ps.

\begin{figure}[H]
\centering
\includegraphics[width=1.0\linewidth]{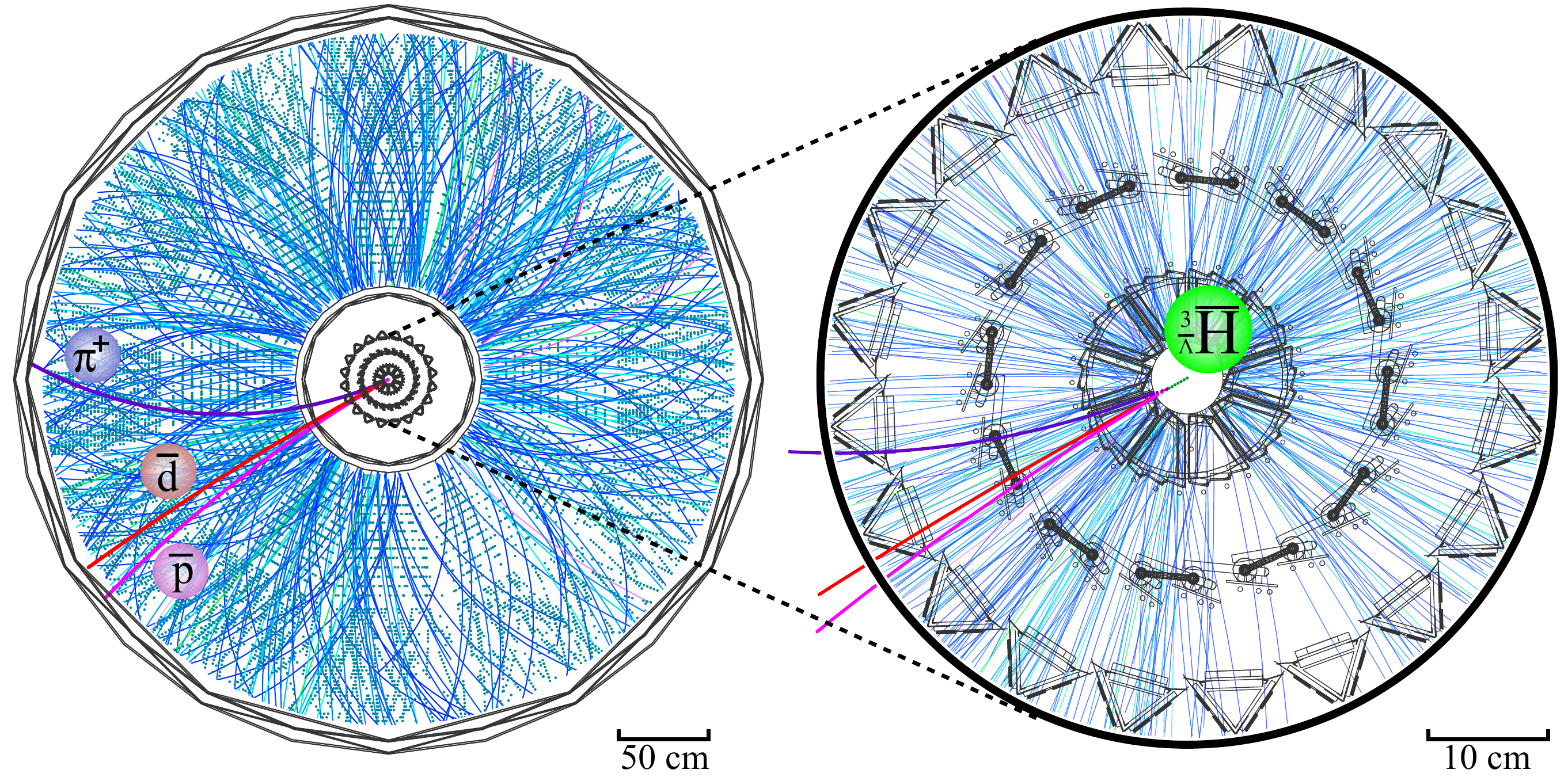}
\caption{\textbf{A typical $\rm^3_{\bar{\Lambda}}\overline{H}$ 3-body decay in the detectors}. The left side shows a less magnified view of the Solenoidal Tracker At RHIC (Relativistic Heavy Ion Collider) detector with the beam axis normal to the page, including a projected view of the large number of tracks detected by the Time Projection Chamber in a typical gold-gold collision. The right side shows a magnified view of the four cylindrical layers of the Heavy Flavor Tracker located at the center of the Time Projection Chamber. The bold red, pink and violet curves represent the trajectories of the $\bar{d}$, $\bar{p}$ and $\pi^+$ decay daughters, respectively. The reconstructed decay daughters can be traced back to the decay vertex, where the $\rm^3_{\bar{\Lambda}}\overline{H}$ decays after flying a distance on the order of centimeters, as shown by the dashed green curve starting at the center of the Heavy Flavor Tracker. }
\label{eventdisplay}
\end{figure}

The hypernucleus $\rm^3_{\Lambda}H$ is reconstructed through its mesonic decay channels $\rm^3_{\Lambda}H  \rightarrow {^3}He + \pi^-$ (2-body decay) and $\rm^3_{\Lambda}H \rightarrow$ $ d + p + \pi^-$ (3-body decay). Fig.~\ref{eventdisplay} depicts a typical event in which a $\rm^3_{\bar{\Lambda}}\overline{H}$ candidate decays to $\bar{d} + \bar{p} + \pi^+$ in the STAR HFT and TPC. The $\rm^3_{\bar{\Lambda}}\overline{H}$ candidate is produced at the primary vertex of a gold-gold collision and remains in flight for a distance on the order of centimeters, as shown by the dashed green curve starting at the center of the right-hand side of the figure, before decaying as depicted by the bold coloured curves. 

\begin{figure}[hbt]
\centering
\includegraphics[width=1.0\linewidth]{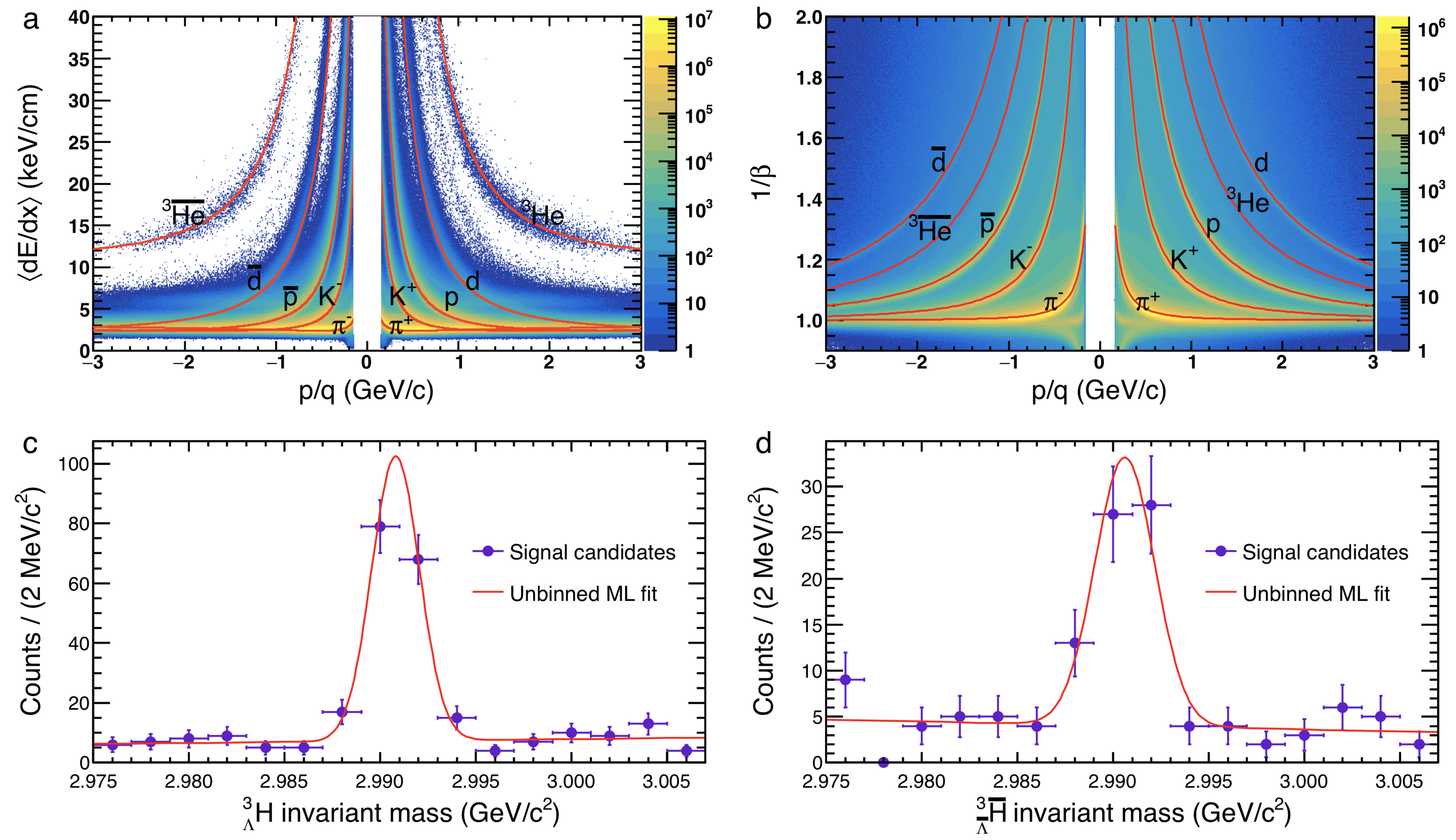}
\caption{\textbf{Particle identification and the invariant mass distributions for $\rm^3_{\Lambda}H$ and $\rm^3_{\bar{\Lambda}}\overline{H}$ reconstruction.} $\langle dE/dx\rangle$ (mean energy loss per unit track length in the gas of the Time Projection Chamber) versus $p/q$ (where $p$ is the momentum and $q$ is the electric charge in units of the elementary charge $e$) is presented in panel a, and $1/\beta$ (where $\beta$ is the speed of a particle in units of the speed of light) versus $p/q$ in panel b. $\langle dE/dx\rangle$ is measured by the Time Projection Chamber and $1/\beta$ is measured by the Time Of Flight detector in conjunction with the Time Projection Chamber. In both cases, the coloured bands show the measured data for each species of charged particle, while the red curves show the expected values. Charged particles are identified by comparing the observed $\langle dE/dx\rangle$ and $1/\beta$ with the expected values. Utilizing both  2-body and 3-body decay channels, the invariant mass distributions of $^3_\Lambda$H and $\rm^3_{\bar{\Lambda}}\overline{H}$ are shown as data points in panels c and d, respectively. The error bars here represent the statistical uncertainties (standard deviations). The red curves represent a fit with a Gaussian function plus a linear background, using the unbinned Maximum Likelihood method.
}
\label{reconstruction}
\end{figure}

Comparisons of the measured $\langle dE/dx\rangle$ and $\beta$ values for each track with their expected values under different mass hypotheses allow decay daughters to be identified. Panel a of Fig.~\ref{reconstruction} presents $\langle dE/dx\rangle$ versus rigidity ($p/q$, where $p$ is the momentum and $q$ is the electric charge in units of the elementary charge $e$), while panel b shows $1/\beta$ versus rigidity. It can be seen that the decay daughter species for $^3_\Lambda$H and $\rm^3_{\bar{\Lambda}}\overline{H}$ are cleanly identified over a wide rigidity range. The helical trajectories of the decay daughter particles can be followed back in time to each secondary decay vertex and used to reconstruct the decay topology of the parent hypernucleus or antihypernucleus. The effects of energy loss (ranging from about 0.2\% for $\pi^{\pm}$ to about 3\% for $^3$He) and TPC field distortion on the measured momenta of the decay daughters are corrected for by data-driven calibration using the world-average $\Lambda$ mass compiled by the PDG {\cite{PDG}}. Due to the high-precision tracking and particle identification capabilities of the STAR experiment, the invariant mass ($\sqrt{(\sum E_i)^2 - (\sum\vec{p}_i)^2}$, where $E_i$ is the energy and $\vec{p}_i$ the momentum of the $i$th decay daughter) of each parent is reconstructed with a low level of background as shown in panels c and d of Fig. \ref{reconstruction}. The background originates from combinatorial contamination and particle misidentification. The significance $S/ \sqrt{S+B}$, where $S$ is signal counts and $B$ is background counts in the invariant mass window $2.986 - 2.996$ GeV$/c^{2}$, is 11.4 for $^3_\Lambda$H and 6.4 for $\rm^3_{\bar{\Lambda}}\overline{H}$. The signal counts from 2-body/3-body decay channels are about 121/35 for $^3_\Lambda$H and 36/21 for $\rm^3_{\bar{\Lambda}}\overline{H}$, respectively. The $^{3}_{\Lambda}$H signal-to-background ratio is close to a factor of 23 better than an earlier measurement from the same experiment using only the TPC{\cite{STAR-H3L-lifetime}}. 

The hypernucleus and antihypernucleus invariant mass distributions reconstructed through 2-body and 3-body decays are each fitted with a Gaussian function plus a straight line, using the unbinned maximum likelihood method. Mass parameters are extracted from the peaks of the invariant mass distributions. Final results are the average of the masses from 2-body and 3-body decays weighted by the reciprocal of the squared statistical uncertainties. The main systematic uncertainty arises from imperfections in the energy loss and field distortion corrections applied to the tracking of decay daughters, estimated to be 0.11 MeV/$c^2$ (37 ppm). Other sources of systematic uncertainty, including those from event selection, track quality cuts, decay topology cuts and fit procedure, are negligible. Accordingly, the measured masses are
$$ m_{^3_{\Lambda}{\rm H}} 
     = 2990.95\pm 0.13 {\rm(stat.)}\pm 0.11 {\rm(syst.)~MeV}/c^2 $$
$$ m_{\rm^3_{\bar{\Lambda}}\overline{H}} 
     = 2990.60\pm 0.28 {\rm(stat.)}\pm 0.11 {\rm(syst.)~MeV}/c^2 $$
The average mass (weighted by the reciprocal of squared statistical uncertainties) for $^3_\Lambda$H and $\rm^3_{\bar{\Lambda}}\overline{H}$ combined is 
\begin{equation}
\label{mass}
m  = 2990.89\pm 0.12 {\rm(stat.)}\pm 0.11 {\rm(syst.)~MeV}/c^2 
\end{equation}

\begin{figure}[H]
\centering
\includegraphics[width=1.0\linewidth]{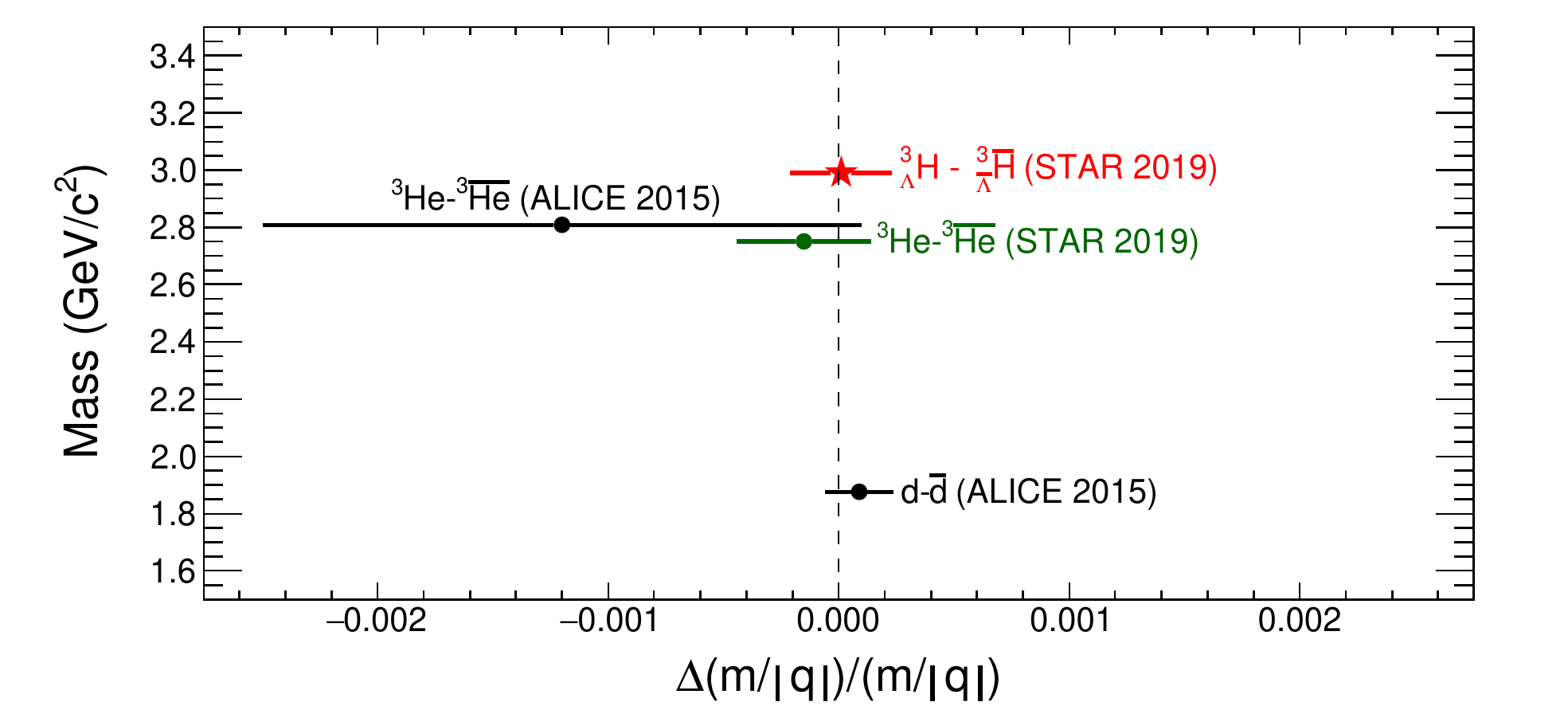}
\caption{\textbf{Measurements of the relative mass-to-charge ratio differences between nuclei and antinuclei.} The current measurement of the relative mass difference ${\Delta m}/m$ between $^3_\Lambda$H and $\rm^3_{\bar{\Lambda}}\overline{H}$ constrained by the existing experimental limits for decay daughters{\cite{massdiff_ALICE}} is shown by the red star marker. The green point is the new $^{3}$He result after applying the constraint provided by the present $^{3}_{\Lambda}$H result. The differences between $d$ and $\bar{d}$ and between $^3$He and $\rm^3\overline{He}$ measured by the ALICE collaboration {\cite{massdiff_ALICE}} are also shown. The two $^3$He - $\rm^3\overline{He}$ points are staggered vertically for visibility. The dotted vertical line at zero is the expectation from CPT invariance. The horizontal error bars represent the sum in quadrature of statistical and systematic uncertainties.
}
\label{massdiff}
\end{figure}

By taking into account the current best limits for the mass differences of $^{3}$He and $d$ reported by ALICE{\cite{massdiff_ALICE}}, the mass differences between $\rm^3_{\Lambda}H$ and $\rm^3_{\bar{\Lambda}}\overline{H}$ are $-2.9 \pm 2.5 {\rm{(stat.)}} \pm 2.8 {\rm{(syst.)}\,MeV/}c^{2}$ and $0.13 \pm 0.63 {\rm{(stat.)}} \pm 0.31 {\rm{(syst.)}\,MeV/}c^{2}$ for 2-body and 3-body decay channels, respectively. The relative mass difference $\Delta m/m$ of 2-body and 3-body decay combined is (see Methods section for details)
\begin{equation}
\label{massdiff_eq2}
\frac{\Delta m}{m} = \frac{m_{{\rm^3_{\Lambda}H}} - m_{{\rm^3_{\bar{\Lambda}}\overline{H}}}}{m} = [\,0.1\pm 2.0 {\rm(stat.)}\pm 1.0 {\rm(syst.)}]\times 10^{-4}
\end{equation}
If we assume CPT symmetry is true for the decay daughters, the relative mass difference between $^3_\Lambda$H and $\rm^3_{\bar{\Lambda}}\overline{H}$ would be $\Delta m/m = [\,1.1\pm 1.0 {\rm(stat.)}\pm 0.5 {\rm(syst.)} ]\times 10^{-4}$.
In addition, by taking the difference between the masses measured in the 2-body and 3-body decay channels of $^{3}_{\Lambda}$H in conjunction with the deuteron masses reported by ALICE{\cite{massdiff_ALICE}}, we can place a new constraint on the relative mass difference between $\rm^{3}He$ and $\rm^{3}\overline{He}$, namely $\rm\Delta m_{{\rm^{3}He}}/m_{{\rm^{3}He}}$ = [-1.5 $\pm$ 2.6 (stat.) $\pm$ 1.2 (syst.)]$\times 10^{-4}$ (see Methods section for details). These results are displayed in Fig.~\ref{massdiff} along with the relative mass-to-charge ratio differences between $d$ and $\bar{d}$ and between $^3$He and $\rm^3\overline{He}$ measured by the ALICE Collaboration{\cite{massdiff_ALICE}}. The mass difference between $^3_\Lambda$H and $\rm^3_{\bar{\Lambda}}\overline{H}$ observed in the present data is consistent with zero, and the precision is an order of magnitude improved over the early data with same mass number{\cite{massdiff_ALICE}}. The current measurement extends the validation of CPT invariance to a nucleus containing a strange quark. 

\begin{figure}[H]
\centering
\includegraphics[width=1.0\linewidth]{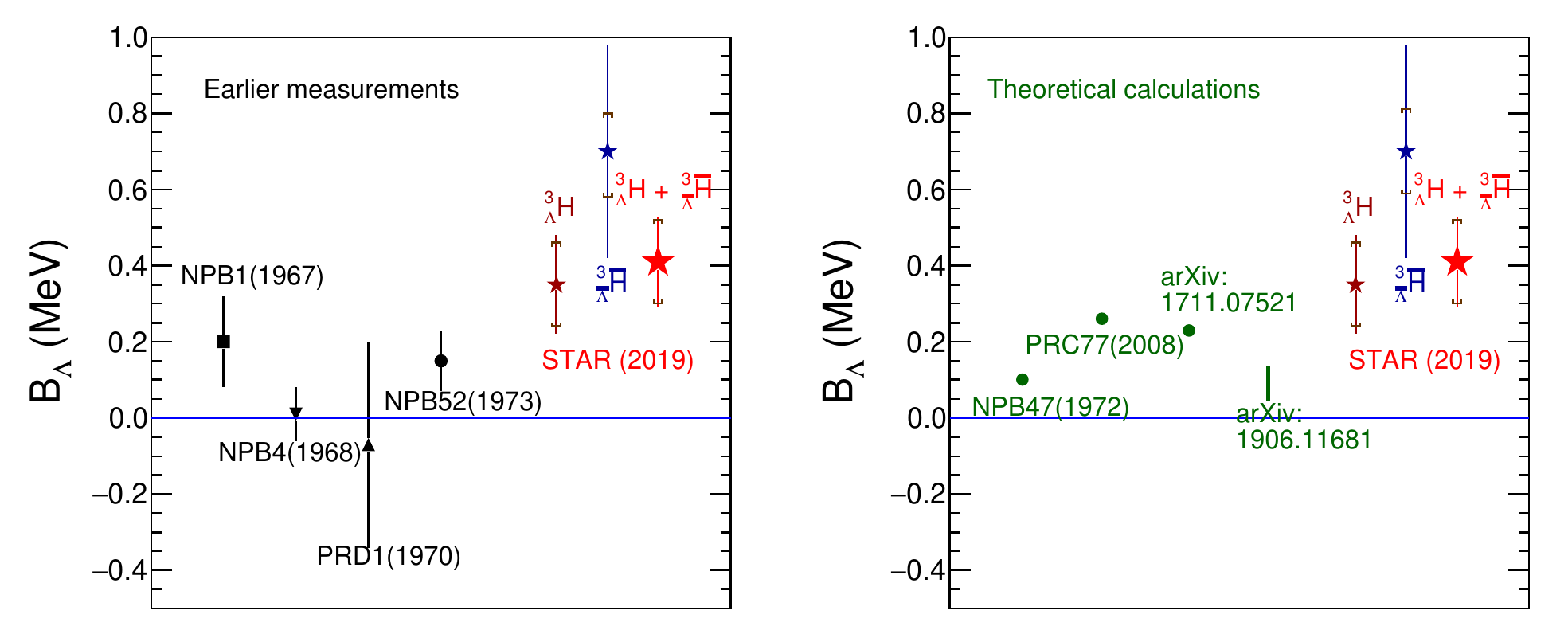}
\caption{\textbf{Measured $\Lambda$ binding energy in the hypertriton compared to earlier results and theoretical calculations.} The black points and their error bars (which are the reported statistical uncertainties) represent $B_\Lambda$ (see text for exact definition) for $^3_\Lambda$H based on earlier data{\cite{B_1967,B_1968,B_1970,B_1973}}. The current STAR measurement plotted here is based on a combination of $^3_\Lambda$H and $\rm^3_{\bar{\Lambda}}\overline{H}$ assuming CPT invariance. Error bars show statistical uncertainties (standard deviations) and caps show systematic errors. The green solid circles and green vertical line in the right panel represent theoretical calculations of $B_\Lambda$ values{\cite{2019_weak, B_1972, QMmodel, AFDMC_2018}}. The horizontal blue lines in both panels indicate a reference energy corresponding to zero binding of the $\Lambda$ hyperon.}
\label{bindingenergy}
\end{figure}

The $\Lambda$ binding energy $B_\Lambda$ for $^3_\Lambda$H and $\rm^3_{\bar{\Lambda}}\overline{H}$ is calculated using the mass measurement shown in equation (\ref{mass}). We obtain
\begin{equation}
\label{bindingE_eq}
B_\Lambda =0.41\pm 0.12 {\rm(stat.)}\pm 0.11{\rm (syst.)~MeV}
\end{equation}
This binding energy is presented in Fig.~\ref{bindingenergy} (left panel) along with earlier measurements{\cite{B_1967,B_1968,B_1970,B_1973}} from nuclear emulsion and helium bubble chamber experiments. The current STAR result differs from zero with a statistical significance of 3.4$\sigma$ and the central value of the current STAR measurement is larger than the commonly used measurement from 1973{\cite{B_1973}}. It has been pointed out in Ref.~{\cite{Achenbach2017}} that for measurements of $B_\Lambda$ for p-shell hypernuclei, there exists a discrepancy in the range of 0.4 to 0.8 MeV between emulsion data and other modern measurements. Whether the effect would be similar in s-shell hypernuclei such as the hypertriton is unclear, but such a discrepancy is much larger than the systematic uncertainty of 0.04 MeV assigned to emulsion measurements{\cite{Davis_2005}}. Until this discrepancy is well understood, an average of the current measurement with early results cannot be reliably carried out.

Theoretical calculations of $B_\Lambda$ for $^3_\Lambda$H are also available (see right panel of Fig.~\ref{bindingenergy}). For example, Dalitz reported the calculation $B_\Lambda = 0.10$ MeV in 1972{\cite{B_1972}}. In recent calculations, $B_\Lambda = 0.262$ MeV was obtained through SU(6) quark model baryon-baryon interactions{\cite{QMmodel}}, and $B_\Lambda$ was calculated to be 0.23 MeV using auxiliary field diffusion Monte Carlo (AFDMC){\cite{AFDMC_2018}}. A span of values ranging from 0.046 MeV to 0.135 MeV was obtained in SU(3) chiral effective field theory{{\cite{2019_weak}}}. The divergence of results among different calculations emphasizes the need for a precise determination of $B_\Lambda$ from experiment. In Ref.{\cite{SL_2019}} a model based on effective field theory is used to extract a scattering length of $13.80^{+3.75}_{-2.03}$ fm from the earlier average value of $\rm 0.13\pm 0.05(stat.)\,MeV$; when applied to our value of $\rm 0.41\pm 0.12(stat.)\,MeV$ it yields a significantly smaller value of $7.90^{+1.71}_{-0.93}$ fm. The larger $B_\Lambda$ and shorter effective scattering length suggest a stronger $YN$ interaction between the $\Lambda$ and the relatively low-density nuclear core of the $^3_\Lambda$H{\cite{stronger_YN}}. This, in certain models, requires SU(3) symmetry breaking and a more repulsive $YN$ interaction at high density, consistent with implications from the range of masses observed for neutron stars{\cite{2019_weak}}.\\

\noindent \textbf{Acknowledgements}

The STAR collaboration acknowledges contributions from Veronica Dexheimer, Fabian Hildenbrand, and Hans-Werner Hammer. We thank the RHIC Operations Group and RCF at BNL, the NERSC Center at LBNL, and the Open Science Grid consortium for providing resources and support.  This work was supported in part by the Office of Nuclear Physics within the U.S. DOE Office of Science, the U.S. National Science Foundation, the Ministry of Education and Science of the Russian Federation, National Natural Science Foundation of China, Chinese Academy of Science, the Ministry of Science and Technology of China and the Chinese Ministry of Education, the National Research Foundation of Korea, Czech Science Foundation and Ministry of Education, Youth and Sports of the Czech Republic, Hungarian National Research, Development and Innovation Office, New National Excellency Programme of the Hungarian Ministry of Human Capacities, Department of Atomic Energy and Department of Science and Technology of the Government of India, the National Science Centre of Poland, the Ministry  of Science, Education and Sports of the Republic of Croatia, RosAtom of Russia and German Bundesministerium fur Bildung, Wissenschaft, Forschung and Technologie (BMBF) and the Helmholtz Association.

\noindent \textbf{Author contributions}

All authors made important contributions to this publication, in one or more of the areas of detector hardware and software, operation of the experiment, acquisition of data, and data analysis.  All STAR collaborations who are authors reviewed and approved the submitted manuscript.

\noindent \textbf{Competing interests}

The authors declare no financial and non-financial competing interests.

\newpage
\noindent \textbf{STAR Collaboration}

\noindent \author{}{
J.~Adam$^{6}$,
L.~Adamczyk$^{2}$,
J.~R.~Adams$^{39}$,
J.~K.~Adkins$^{30}$,
G.~Agakishiev$^{28}$,
M.~M.~Aggarwal$^{40}$,
Z.~Ahammed$^{59}$,
I.~Alekseev$^{3,35}$,
D.~M.~Anderson$^{53}$,
A.~Aparin$^{28}$
% ...
}

\begingroup
\centering
\affil{}{$^{1}$Abilene Christian University, Abilene, Texas   79699}\\
\affil{}{$^{2}$AGH University of Science and Technology, FPACS, Cracow 30-059, Poland}\\
\affil{}{$^{3}$Alikhanov Institute for Theoretical and Experimental Physics NRC "Kurchatov Institute", Moscow 117218, Russia}\\
\affil{}{$^{4}$Argonne National Laboratory, Argonne, Illinois 60439}\\
\affil{}{$^{5}$American University of Cairo, New Cairo 11835, New Cairo, Egypt}\\
\affil{}{$^{6}$Brookhaven National Laboratory, Upton, New York 11973}\\
\affil{}{$^{7}$University of California, Berkeley, California 94720}\\
\affil{}{$^{8}$University of California, Davis, California 95616}\\
\affil{}{$^{9}$University of California, Los Angeles, California 90095}\\
\affil{}{$^{10}$University of California, Riverside, California 92521}\\
\affil{}{$^{11}$Central China Normal University, Wuhan, Hubei 430079 }\\
\affil{}{$^{12}$University of Illinois at Chicago, Chicago, Illinois 60607}\\
\affil{}{$^{13}$Creighton University, Omaha, Nebraska 68178}\\
\affil{}{$^{14}$Czech Technical University in Prague, FNSPE, Prague 115 19, Czech Republic}\\
\affil{}{$^{15}$Technische Universit\"at Darmstadt, Darmstadt 64289, Germany}\\
\affil{}{$^{16}$ELTE E\"otv\"os Lor\'and University, Budapest, Hungary H-1117}\\
\affil{}{$^{17}$Frankfurt Institute for Advanced Studies FIAS, Frankfurt 60438, Germany}\\
\affil{}{$^{18}$Fudan University, Shanghai, 200433 }\\
\affil{}{$^{19}$University of Heidelberg, Heidelberg 69120, Germany }\\
\affil{}{$^{20}$University of Houston, Houston, Texas 77204}\\
\affil{}{$^{21}$Huzhou University, Huzhou, Zhejiang  313000}\\
\affil{}{$^{22}$Indian Institute of Science Education and Research (IISER), Berhampur 760010 , India}\\
\affil{}{$^{23}$Indian Institute of Science Education and Research (IISER) Tirupati, Tirupati 517507, India}\\
\affil{}{$^{24}$Indian Institute Technology, Patna, Bihar 801106, India}\\
\affil{}{$^{25}$Indiana University, Bloomington, Indiana 47408}\\
\affil{}{$^{26}$Institute of Modern Physics, Chinese Academy of Sciences, Lanzhou, Gansu 730000 }\\
\affil{}{$^{27}$University of Jammu, Jammu 180001, India}\\
\affil{}{$^{28}$Joint Institute for Nuclear Research, Dubna 141 980, Russia}\\
\affil{}{$^{29}$Kent State University, Kent, Ohio 44242}\\
\affil{}{$^{30}$University of Kentucky, Lexington, Kentucky 40506-0055}\\
\affil{}{$^{31}$Lawrence Berkeley National Laboratory, Berkeley, California 94720}\\
\affil{}{$^{32}$Lehigh University, Bethlehem, Pennsylvania 18015}\\
\affil{}{$^{33}$Max-Planck-Institut f\"ur Physik, Munich 80805, Germany}\\
\affil{}{$^{34}$Michigan State University, East Lansing, Michigan 48824}\\
\affil{}{$^{35}$National Research Nuclear University MEPhI, Moscow 115409, Russia}\\
\affil{}{$^{36}$National Institute of Science Education and Research, HBNI, Jatni 752050, India}\\
\affil{}{$^{37}$National Cheng Kung University, Tainan 70101 }\\
\affil{}{$^{38}$Nuclear Physics Institute of the CAS, Rez 250 68, Czech Republic}\\
\affil{}{$^{39}$Ohio State University, Columbus, Ohio 43210}\\
\affil{}{$^{40}$Panjab University, Chandigarh 160014, India}\\
\affil{}{$^{41}$Pennsylvania State University, University Park, Pennsylvania 16802}\\
\affil{}{$^{42}$NRC "Kurchatov Institute", Institute of High Energy Physics, Protvino 142281, Russia}\\
\affil{}{$^{43}$Purdue University, West Lafayette, Indiana 47907}\\
\affil{}{$^{44}$Rice University, Houston, Texas 77251}\\
\affil{}{$^{45}$Rutgers University, Piscataway, New Jersey 08854}\\
\affil{}{$^{46}$Universidade de S\~ao Paulo, S\~ao Paulo, Brazil 05314-970}\\
\affil{}{$^{47}$University of Science and Technology of China, Hefei, Anhui 230026}\\
\affil{}{$^{48}$Shandong University, Qingdao, Shandong 266237}\\
\affil{}{$^{49}$Shanghai Institute of Applied Physics, Chinese Academy of Sciences, Shanghai 201800}\\
\affil{}{$^{50}$Southern Connecticut State University, New Haven, Connecticut 06515}\\
\affil{}{$^{51}$State University of New York, Stony Brook, New York 11794}\\
\affil{}{$^{52}$Temple University, Philadelphia, Pennsylvania 19122}\\
\affil{}{$^{53}$Texas A\&M University, College Station, Texas 77843}\\
\affil{}{$^{54}$University of Texas, Austin, Texas 78712}\\
\affil{}{$^{55}$Tsinghua University, Beijing 100084}\\
\affil{}{$^{56}$University of Tsukuba, Tsukuba, Ibaraki 305-8571, Japan}\\
\affil{}{$^{57}$United States Naval Academy, Annapolis, Maryland 21402}\\
\affil{}{$^{58}$Valparaiso University, Valparaiso, Indiana 46383}\\
\affil{}{$^{59}$Variable Energy Cyclotron Centre, Kolkata 700064, India}\\
\affil{}{$^{60}$Warsaw University of Technology, Warsaw 00-661, Poland}\\
\affil{}{$^{61}$Wayne State University, Detroit, Michigan 48201}\\
\affil{}{$^{62}$Yale University, New Haven, Connecticut 06520}\\
\endgroup

\newpage
\noindent {\bf Methods}

In this section, we provide more details on how a possible CPT violation of $^{3}$He and $d$ would impact the present measurement of mass difference between $\rm ^3_{\Lambda}H$ and $\rm ^3_{\bar{\Lambda}}\overline{H}$. The reconstruction of $\rm ^3_{\Lambda}H$ mass from its 2-body decay kinematics: 
\begin{equation}
\begin{aligned}
\label{2bodydecay}
m^{2} = \left(E_{\pi^{-}} + E_{\rm{^{3}He}}\right)^{2} - \left(\overrightarrow{p}_{\pi^{-}} + \overrightarrow{p}_{\rm{^{3}He}}\right)^{2} 
\end{aligned}
\end{equation}
\noindent from its 3-body decay kinematics:
\begin{equation}
\begin{aligned}
\label{3bodydecay}
m^{2} = \left(E_{\pi^{-}} + E_{p} + E_{d}\right)^{2} - \left(\overrightarrow{p}_{\pi^{-}} + \overrightarrow{p}_{p} + \overrightarrow{p}_{d}\right)^{2} 
\end{aligned}
\end{equation}
where $E_{\pi^{-}}$, $E_{p}$, $E_{d}$, and $E_{\rm{^{3}He}}$ are the total energies of decay daughters of $\rm ^3_{\Lambda}H$, $\overrightarrow{p}_{\pi^{-}}$, $\overrightarrow{p}_{p}$, $\overrightarrow{p}_{d}$, and $\overrightarrow{p}_{\rm{^{3}He}}$ are the momentum vectors of decay daughters of $\rm ^3_{\Lambda}H$.

If we assume a possible CPT violation among the $^{3}$He and $d$ daughters and that neither decay process produces or reduces any CPT asymmetry, the measured mass difference between $\rm ^3_{\Lambda}H$ and $\rm ^3_{\bar{\Lambda}}\overline{H}$ depends on the CPT violations in $\rm ^3_{\Lambda}H$, $^{3}$He and $d$.
\begin{equation}
\begin{aligned}
\label{new-2body}
\Delta m^{\rm{2-body}} &= 
\Delta m_{{\rm ^3_{\Lambda}H}}  - 
\Delta m_{{\rm ^{3}He}} \left( 1+\sqrt{\frac{m^{2}_{\pi^{-}}  +  \overrightarrow{p}^{2}_{\pi^{+}}}    {m^{2}_{\rm ^{3}He}  +  \overrightarrow{p}^{2}_{\rm ^{3}\overline{He}}}} \right) \frac{m_{\rm ^{3}He}}{m}
\approx 
\Delta m_{{\rm ^3_{\Lambda}H}}  - 1.01\Delta m_{{\rm ^{3}He}}
\end{aligned}
\end{equation}

\begin{equation}
\begin{aligned}
\label{new-3body}
\Delta m^{\rm{3-body}} &= 
\Delta m_{{\rm ^3_{\Lambda}H}}  - 
\Delta m_{{d}} \left( 1+\sqrt{\frac{m^{2}_{\pi^{-}}  +  \overrightarrow{p}^{2}_{\pi^{+}}}{m^{2}_{d}  +  \overrightarrow{p}^{2}_{\bar{d}}}} 
+ 
\sqrt{\frac{m^{2}_{p}  +  \overrightarrow{p}^{2}_{\bar{p}}}    {m^{2}_{d}  +  \overrightarrow{p}^{2}_{\bar{d}}}} \right)  \frac{m_{d}}{m} 
\approx 
\Delta m_{{\rm ^3_{\Lambda}H}}  - 1.00\Delta m_{{d}}
\end{aligned}
\end{equation}
where $\rm \Delta m_{{\rm ^{3}He}} = m_{^{3}He} - m_{^{3}\overline{He}}$ and $\Delta m_d = m_d - m_{\bar{d}}$ are the hypothesized mass symmetry violations for $\rm ^{3}He$ and $d$. The mass difference of $\pi^{\pm}$ presented in PDG{\cite{PDG}} is negligible at the precision reported in this letter. $\overrightarrow{p}_{\pi^{+}}$, $\overrightarrow{p}_{\bar{p}}$, $\overrightarrow{p}_{\bar{d}}$, and $\overrightarrow{p}_{^{3}\rm{\overline{He}}}$ are the momentum of $\pi^{+}$, $\bar{p}$, $\bar{d}$, and $^{3}\rm{\overline{He}}$ when $\rm ^3_{\bar{\Lambda}}\overline{H}$ decays and taken from experiment data. We have separated the mass measurements in 2-body and 3-body decays: $\Delta m^{\rm{2-body}} = 0.56 \pm 0.37 \rm{(stat.)} \pm 0.16 \rm{(syst.)}$ MeV/$c^{2}$, $\Delta m^{\rm{3-body}} = -0.04 \pm 0.62 \rm{(stat.)} \pm 0.16 \rm{(syst.)}$ MeV/$c^{2}$.

According to equations (\ref{new-2body}) and (\ref{new-3body}), the limits of mass differences on $^{3}$He and $d$ from ALICE {\cite{massdiff_ALICE}} in conjunction with our measurements provide the mass differences:
 $$\Delta m_{{\rm ^3_{\Lambda}H}} = -2.9 \pm 2.5 {\rm{(stat.)}} \pm 2.8 {\rm{(syst.)}\,MeV/}c^{2}$$
 $$\Delta m_{{\rm ^3_{\Lambda}H}} = 0.13 \pm 0.63 {\rm{(stat.)}} \pm 0.31 {\rm{(syst.)}\,MeV/}c^{2}$$
  for 2-body and 3-body decay channels, respectively. The weighted mass difference (weighted by the reciprocal of the squared sum in quadrature of statistical and systematic uncertainties) of 2-body and 3-body is
  $$\Delta m_{{\rm ^3_{\Lambda}H}} = 0.03 \pm 0.62 {\rm{(stat.)}} \pm 0.31 {\rm{(syst.)}\,MeV/}c^{2}$$
Accordingly the relative mass difference between $\rm ^3_{\Lambda}H$ and $\rm ^3_{\bar{\Lambda}}\overline{H}$ is
$$ \frac{\Delta m}{m} = \frac{m_{{\rm^3_{\Lambda}H}} - m_{{\rm^3_{\bar{\Lambda}}\overline{H}}}}{m} = [\,0.1\pm 2.0 {\rm(stat.)}\pm 1.0 {\rm(syst.)} ]\times 10^{-4} $$
where $m$ is from equation (\ref{mass}).

According to equations (\ref{new-2body}) and (\ref{new-3body}), any measured difference between 2-body and 3-body decays $\left(\Delta m^{\rm{3-body}} - \Delta m^{\rm{2-body}}\right)$ is due to the CPT violations in the daughter sectors:
\begin{equation}
\begin{aligned}
\label{newmassdiff}
\Delta m_{\rm{^{3}He}} = 0.99\left(\Delta m^{\rm{3-body}} - \Delta m^{\rm{2-body}}\right) + 0.99\Delta m_{{d}}
\end{aligned}
\end{equation}
These present measurements in conjunction with the deuteron reported by ALICE{\cite{massdiff_ALICE}}, place a new constraint on the mass difference between $\rm ^{3}He$ and $\rm ^{3}\overline{He}$, namely
$\Delta m_{\rm{^{3}He}} = -0.43 \pm 0.72 {\rm{(stat.)}} \pm 0.34 {\rm{(syst.)\,MeV}}/c^2$ and $\Delta m_{\rm{^{3}He}}/m_{\rm{^{3}He}} = (-1.5 \pm 2.6 {\rm{(stat.)}} \pm 1.2 {\rm{(syst.)}})\times 10^{-4}$.

~\\
~\\
\noindent {\bf Data availability}

All raw data for this study were collected by the STAR detector at Brookhaven National Laboratory. Data tables for the results reported in this letter are publicly available on the STAR website 
(\href{https://drupal.star.bnl.gov/STAR/publications/precise-measurement-mass-difference-and-binding-energy-hypertriton-and-antihypertrito-0}{https://drupal.star.bnl.gov/STAR/publications/precise-measurement-mass-difference-and-binding-energy-hypertriton-and-antihypertrito-0}) or from the corresponding author P. Liu, J. H. Chen upon reasonable request.

\end{document}